\documentstyle[11pt,newpasp,twoside,epsf]{article}

\begin{document}

\title{A Pilot Study of the Tully-Fisher Relation in Cl0024+1654 at z=0.4}

\author{Anne J. Metevier and David C. Koo}
\affil{Department of Astronomy and Astrophysics, University of California,
 Santa Cruz, CA 95064 USA}

\author{Luc Simard}
\affil{Steward Observatory,
University of Arizona,
Tucson, AZ, 85721 USA}

\begin{abstract}

We present the first determination of the Tully-Fisher relation (TFR) for a distant galaxy
cluster.  We have measured internal kinematics 
of seven members of Cl0024+1654 from spatially resolved emission
observed in 2D Keck (LRIS) spectra.  Measurements of disk structural parameters were made from HST (WFPC2) images. While we do not find evidence for a change in 
the slope or zeropoint of the
cluster TFR, we do find that there was more scatter in the past.
We intend to expand this study with kinematic measurements from fifteen more
Cl0024 members.

\end{abstract}

\section{Motivation}

%	Should I highlight the fact that clusters are ideal labs from which 
%to study galaxy evolution here?

	Since Butcher and Oemler's (1978, hereafter BO78; 1984) pioneering studies of the color
evolution of galaxies in clusters, many physical mechanisms have been proposed 
as causes for this observed evolution.  Viable processes 
can be divided into two groups: those involving interactions between the ICM 
and cluster galaxies (e.g., ram pressure stripping) and those resulting from 
the high density of galaxies in the cluster environment (e.g., galaxy 
harassment).  All of these processes involve disks, gas, and star 
formation.  While many recent cluster studies have concentrated on the 
early-type population of galaxies in clusters (e.g., fundamental plane study by Kelson et al. 2000), we would like to focus on disk systems.  To begin our examination of cluster disks, we have undertaken a pilot 
study of the Tully-Fisher relation (TFR) in Cl0024.  

\section{Target and Dataset}

Cl0024 is an excellent candidate cluster for the study of disk galaxies.  
As one of the two clusters originally studied by BO78, Cl0024 has a high 
fraction of blue galaxies. It is one of the richest clusters known, making
it a good candidate for a study of intracluster effects.  Furthermore, 
Cl0024 is one of the 
best-studied clusters.  Recent studies of Cl0024 members include a fundamental
plane investigation (van Dokkum \& Franx 1996), a look at strong emission line 
galaxies (Koo et al. 1997), and a study of chemical abundances
(Kobulnicky \& Zaritsky 1999).  Morphological classification
of Cl0024 galaxies (Smail et al. 1997) has indicated a 
significant fraction of spirals in the cluster, in agreement with BO78.  Furthermore, Cl0024 exhibits a spectacular gravitational lens 
system with a counter-arc, first discovered by Koo (1987). This system has prompted cluster mass measurements via lensing models
(e.g., Broadhurst et al. 2000) which may be compared with X-ray 
mass estimates, e.g., Soucail et al. (2000).

We have undertaken a Keck (LRIS) spectral survey of Cl0024 (see Metevier et al., in these proceedings, for more details).  The wavelength range of this survey covers restframe [OII]
through H$\alpha$ for the cluster, and the spectral and spatial resolution are higher than those of other recent surveys of this cluster (Dressler et al. 1999, Czoske et al. 2001).  Furthermore, our slits are aligned with galaxy major axes.  These spectra provide us with spatially resolved emission for several cluster members.  To complement our spectra, we have obtained WFPC2 images of Cl0024 in two filters (25 ksec in F450W, 20 
ksec in F814W) from the HST archive.  The high resolution of these images is 
necessary for deriving disk size scales, inclinations, colors, and structural parameters crucial to Tully-Fisher work.

\section{Cl0024 Tully-Fisher Relation}

	The Tully-Fisher relation (TFR) is a scaling relation between 
	galaxy absolute magnitudes and the terminal 
	velocities of their rotation curves and serves a tool for studying galaxy 
	mass-to-light ratios (M/L).  In the field, the TFR has been 
	measured out to z=1 (e.g., 
	Vogt et al. 1996, 1997; Simard \& Pritchet 1998).  However, the cluster TFR
	has only been well-studied out to z=0.1 (e.g., Dale et al. 
	 2000).  Up to now, a key question has been whether or not emission 
	in distant cluster disk systems is spatially extended enough 
	for rotation curve analysis and Tully-Fisher measurements.

%	Galaxies in clusters are known to be affected by different physical
%	processes than field galaxies, but it is yet unknown how these 
%	processes affect the cluster TFR.  For example, a change
%	in the zeropoint of the cluster TFR at intermediate redshift 
%	toward higher magnitudes (lower M/L) might
%	indicate that 
%	cluster galaxies were experiencing starbursts in the past.  Studies of the
%	cluster TFR at z$>$0.1 are necessary for
%	understanding how the physical processes that affect cluster galaxy
%	evolution in turn affect the cluster Tully-Fisher relation.

%\subsection{Cl0024 Rotation Curve Measurements and TFR}

We have measured the rotation curves of a sample of 7 galaxies in the Cl0024
field.  Each is a cluster member and lies within the 
HST images.  Furthermore, all 7 galaxies exhibit emission
which can be measured out to a radius $\ge 1\arcsec$ and are objects for which
our spectral slits were aligned with the galaxy major axes.
We fit a Gaussian to the following emission lines: H$\alpha$, [NII], [OIII], 
H$\beta$, and [OII].  Arctan fits (Willick 1999) were then applied to the resulting 
rotation curves (see Figure 1$a$ for an example). 

In the future, we will improve these fits by modelling and correcting for
the effects of seeing, slit widths, and slit positioning effects.  We will also add internal extinction corrections.  Furthermore, we are studying cross-correlation routines and 2D modelling as alternatives to our Gaussian emission line fits.

\begin{figure}
\plotone{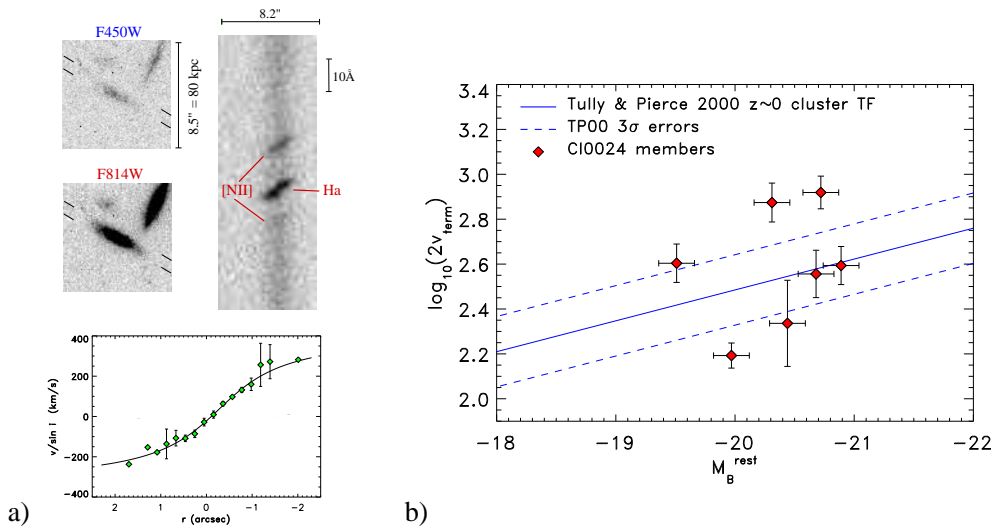}
\caption{$a)$ WFPC2 images, 2D LRIS spectrum, and rotation curve for a Cl0024 member.  Rotation curve errors do not include uncertainties in inclination. $b)$ Cl0024 Tully-Fisher measurements show more scatter than the local relation derived by Tully \& Pierce (2000).}
\end{figure}

In Figure 1$b$, our Tully-Fisher measurements of Cl0024 members are compared to the z$\simeq$0 cluster TFR derived by 
Tully \& Pierce (2000).  While we find no evidence within the presently very small sample for any large changes in the slope or
zeropoint of the cluster TFR with redshift, we do see an increase in scatter
in the relation at z=0.4.  This result is in agreement with the results of 
Dale et al. (2000), who found a $50\%$ higher scatter in the z=0.08 cluster
TFR than in the 0.02$<$z$<$0.06 relation.  

We do not have enough data to make stronger statements about the
evolution of the cluster TFR.  We can, however, explain the higher scatter in the cluster TFR in the past.  A galaxy with a high M/L (scattered high on the TFR) in the 
past may result from the fact that its
gas was stripped, e.g., via ram pressure stripping, and hence star formation was quenched.  By the present epoch,
the gas in such a galaxy may be entirely stripped such that it no longer 
exhibits emission (i.e., we no longer include it in a Tully-Fisher sample).  On the other hand,
galaxies scattered low on the TFR (low M/L) in the past may result from enhanced starburst activity.  Such galaxies may not be experiencing such dramatic star
formation in the present day and therefore their scatter on the diagram is dampened.

\section{Results and Future Work}

The three key results of our pilot study are the following:
{\bf 1)} We have shown that some intermediate-$z$ cluster spirals are large enough and have strong enough emission lines to do rotation curve analysis.
{\bf 2)} Our data already yield 7 rotation curves in the HST-imaged region of Cl0024.
	We have 15 rotation curve candidates outside this region which will
	expand our sample.  While small, the sample is comparable in
	size to that studied by, e.g., Vogt et al. (1996).  Similar quality data for 5--10 more clusters would tightly constrain the 
	intermediate-redshift cluster TFR.
{\bf 3)} The scatter in our preliminary measurement of the cluster TFR 
	at z=0.4 is larger than that measured at z=0.  Within an admittedly sparse sample, we find no evidence for a 
	change in TF slope or zeropoint.

We will expand our study in the future by examining emission line
diagnostics of physical processes affecting Cl0024 disk systems.  Our spectral
resolution and wavelength range will provide 
measurements of star formation rates from H$\alpha$, of
metallicities via the R23 index (from [OII], [OIII], and H$\beta$), and of the ionization state of the emission gas via 
line ratios of H$\alpha$, [NII], [SII], [OIII], and H$\beta$.  Evidence
for shock ionization might be expected from interactions between cluster
galaxies and the ICM.  We also add photometric 
measurements from the archival WFPC2 images of this cluster and structural 
models of the cluster galaxies.  The outcome of these efforts will provide the most
complete examination of the processes affecting disk systems in a single distant cluster.

\end{document}